\documentclass[preprint,12pt]{elsarticle}
\usepackage{graphicx}
\usepackage{amsmath}   
\usepackage{etoolbox}
\usepackage{amssymb}
\usepackage{epsfig}
\usepackage{lineno, hyperref}
\usepackage{url}
\usepackage{amsthm}
\usepackage[export]{adjustbox}
\usepackage[font=scriptsize,labelfont=bf,skip=5pt]{caption} 
\usepackage[belowskip=-25pt,aboveskip=2pt]{caption}
\usepackage{subfig}
\usepackage{xpatch}
\usepackage{hyperref}
\hypersetup{
    colorlinks=true,
    linkcolor=blue,
    filecolor=magenta,      
    urlcolor=cyan,
    pdftitle={ATLAS RPC Hermetic Sealing},
    pdfpagemode=FullScreen,
    }

\def\pmbanner{}

\begin{document}
\begin{frontmatter}

\title{ \pmbanner {Performance studies of thin gas gap Resistive Plate Chamber prototypes with low Global Warming Potential gases for the ANUBIS experiment}}
\author[]{Aashaq Shah\corref{cor}}
\ead{aashaq.shah@cern.ch}
\cortext[cor]{Corresponding author}
\author[]{Thomas Adolphus}
\author[]{Yingchang Zhang}
\author[]{Oleg Brandt}
\address{Cavendish Laboratory, University of Cambridge, Cambridge, United Kingdom}


\begin{abstract}
Resistive Plate Chambers (RPCs) have traditionally operated with high Global Warming Potential (GWP) gas mixtures, adding to the environmental footprint of large-scale physics experiments. In response, efforts are underway to explore environmentally friendly alternatives as a long-term solution and low-GWP as a feasible short- to medium-term replacement for standard RPC gases. This study tests a few mixtures in 50 cm $\times$ 50 cm, 1 mm single-gap High-Pressure Laminate (HPL) RPC prototypes, as part of ongoing efforts for the ANUBIS experiment, which will operate with a 9.8 m$^{3}$ active gas volume. Measurements of performance metrics, including current and efficiency, are conducted with both standard and modified mixtures to assess their viability in sustaining detector performance. The results are also relevant for large RPC systems in other experiments at the LHC, such as ATLAS and CMS, as well as in applications beyond the LHC, supporting a shift toward environmentally sustainable gas mixtures in particle physics detectors.
\end{abstract}

\end{frontmatter}

\section{Introduction}
\begin{figure}[ht]
    \vspace{0.3cm}
    \centering
    \includegraphics[width=5.5cm, height=5.0cm]{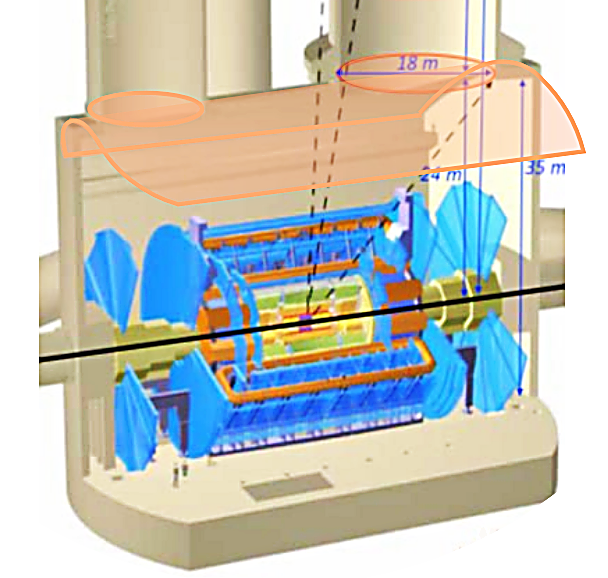}
    \vspace{5pt}
    \caption{The layout of the underground cavern at LHC Point 1 featuring the ATLAS experiment~\cite{Aad:1129811}. The large area highlighted in orange illustrates the ceiling of the ATLAS cavern where ANUBIS tracking layers will be installed. The configuration also includes two disks covering the access shafts which are an integral part of the ANUBIS experiment.}
    \label{fig:anubis_project_layout}
    \vspace{20pt}
\end{figure}

ANUBIS~\cite{Bauer:2019vqk, Shah:2024fpl}, is a proposed Long-Lived Particle (LLP) search experiment at Point 1 of the LHC at CERN. It aims to detect LLP signatures that are predicted in many Beyond the Standard Model (BSM) theories by instrumenting the ceiling of the ATLAS experiment with detector layers to track charged particles, creating an unprecedently large active decay volume extending from the outer layers of ATLAS to the cavern ceilings. To achieve this, the project requires detector technology that can achieve a time resolution better than 0.5 ns, a hit efficiency of approximately 98\% per layer, and a spatial resolution of $\mathcal{O}$(cm). Resistive Plate Chamber (RPC) technology~\cite{Santonico:1981sc} meets all these requirements~\cite{Bauer:2019vqk, CATTANI2012S6} and offers a cost-effective solution for large-area instrumentation, making it a natural choice for ANUBIS. The proposed layout includes RPCs covering a total detector area of approximately 9800~m$^2$, as illustrated in Figure~\ref{fig:anubis_project_layout}. Moreover, RPCs are well suited to ANUBIS due to their demonstrated capability in detecting charged particles -- particularly muons, which are expected to be the dominant decay products of LLPs.

High-Pressure Laminate (HPL) RPCs are well suited for ANUBIS, having already demonstrated reliable performance in the muon systems of several LHC experiments, including ALICE~\cite{Ferretti:2018fga}, ATLAS~\cite{Boscherini:2850856}, and CMS~\cite{Shopova:2856419}. These RPCs typically operate with a standard gas mixture composed of 90--95\% Freon (C$_2$H$_2$F$_4$), also known as Tetrafluoroethane or R-134a\footnote{For brevity, referred to as Freon throughout this work, especially in figure captions}, 4.5--10\% isobutane (i-C$_4$H$_{10}$), 0.3\% sulphur hexafluoride (SF$_6$), and 7000--8000 ppm water vapour. However, both Freon and SF$_6$ are potent greenhouse gases, with Global Warming Potentials (GWPs) of 1,430 and 22,800, respectively~\cite{Solomon:IPCC2007}. To align with European strategies aimed at minimizing the use of high-GWP fluorinated gases~\cite{eu03}, LHC collaborations such as ATLAS and CMS are pursuing two key approaches: a short-term strategy to reduce high-GWP gases like Freon by incorporating CO$_2$ -- recently used by ATLAS -- and other low-GWP gases into gas mixtures, and a longer-term effort, also adopted by ANUBIS, to explore alternative environmentally friendly gases for RPCs.  This shift is particularly relevant, as during LHC Run 2 at CERN, RPCs contributed approximately 87\% of the total greenhouse gas emissions largely due to gas leaks at the detector level \cite{Rigoletti:2875180}.

For an alternative gas mixture to be viable, it must satisfy several key criteria: it should be non-flammable and non-toxic to ensure operational safety, maintain long-term stability, and generate sufficient charge in the gas gap to be reliably detected by the front-end electronics. Additionally, the mixture should not adversely affect the detector's lifetime or compromise its rate capability.

In this context, prototype HPL RPCs have been developed specifically to study new and environmentally friendly gas mixtures for their operation in the ANUBIS experiment. Initial tests with these prototypes are conducted using standard gas mixtures to establish a performance baseline, followed by preliminary measurements with low-GWP alternatives to assess the feasibility of environmentally friendly gas mixtures.  While the performance of 2~mm HPL RPCs has been well documented in the literature -- for example, in Refs.~\cite{Rigoletti:2875180, Proto:2024bqw, RPCECOGasGIF:2024ilv} -- our focus on RPC with 1~mm gas gaps addresses a relatively less explored area. This study aims to provide additional insights into the capabilities and operational characteristics of this narrower-gap RPC technology.

\section{Experimental Setup}  
The measurements reported in this paper have been taken using prototype HPL RPCs with a setup comprising a scintillator system for the trigger, a data acquisition (DAQ) setup, and a gas mixing unit. Each of these components is briefly described in the following sections.  

\subsection{Prototype RPC Fabrication}  
\begin{figure}[ht]
\centering
\includegraphics[width=4cm, height= 3cm]{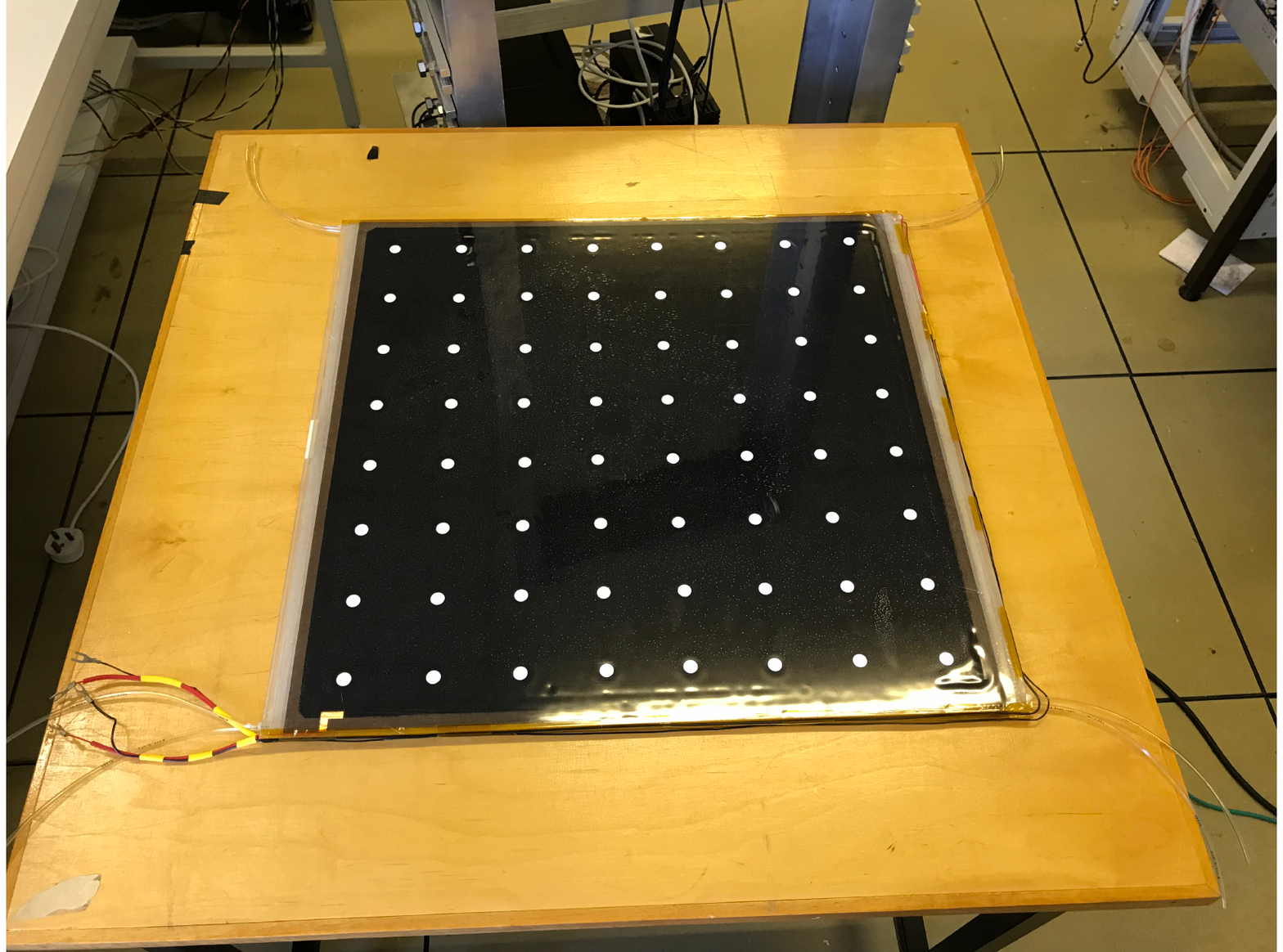}
\includegraphics[width=4cm, height= 3cm]{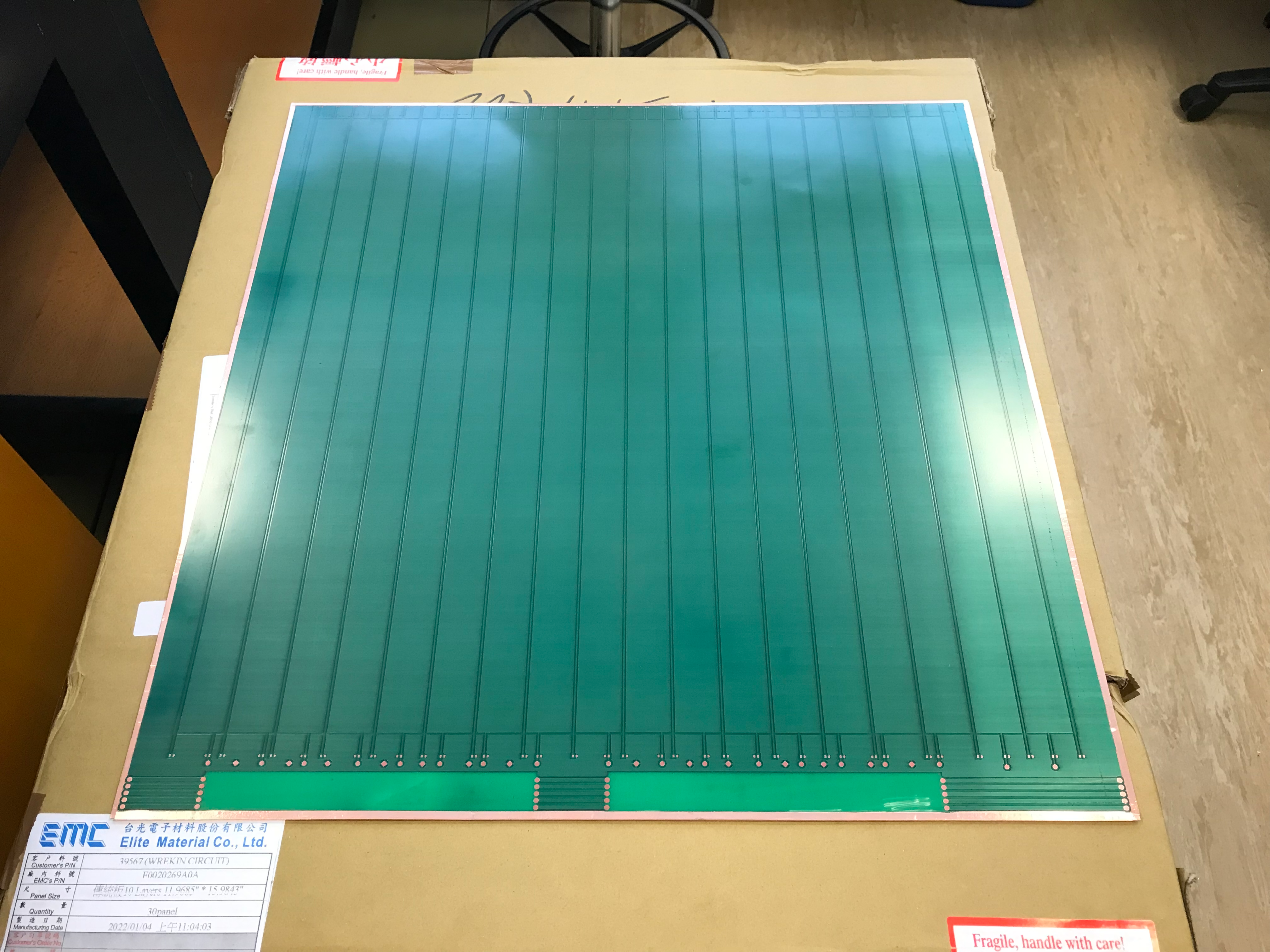}
\caption{(left) An HPL gas gap with dimensions of 50 cm $\times$ 50 cm from GTE. (right) One of the strip panels with a total of 18, 2.5 cm wide copper strips, each coated with a thin (a few microns) insulating layer, visible as the green coating, to ensure surface passivation.} 
\label{fig:Gas_gaps_stripPanel}
\vspace{0.5cm}
\end{figure}

The prototype RPCs have been fabricated using High-Pressure Laminate -- commonly referred to as Bakelite~\cite{Ganai:2015cya} -- electrodes with a thickness of 1.2~mm, supplied by Teknemika to General Tecnica Engineering (GTE) in Italy. The gas gaps, shown in Figure~\ref{fig:Gas_gaps_stripPanel} (left), with dimensions of 50~cm~$\times$~50~cm and a gap thickness of 1~mm, are manufactured by GTE and later procured by the University of Cambridge. The bulk resistivity of the Bakelite electrodes, measured at 20$^{\circ}$C and 50\% relative humidity, typically falls within the range of $1\text{--}5 \times 10^{10}\,\Omega\cdot\text{cm}$~\cite{Rigoletti:2875180}. However, due to variations in production and environmental sensitivity of Bakelite, values up to $10^{11}\,\Omega\cdot\text{cm}$ have occasionally been observed. This upper bound reflects the maximum resistivity encountered during quality control and characterisation, and is used to define acceptance thresholds in electrode selection. The surface resistivity of the graphite coating is approximately $350 \pm 100\,\text{k}\Omega/\Box$, and the pitch between the 1~mm gas gap spacers is about 7~cm in the transverse directions.

\begin{figure}[ht]
\centering
\includegraphics[width=8.7cm, height= 5.3cm]{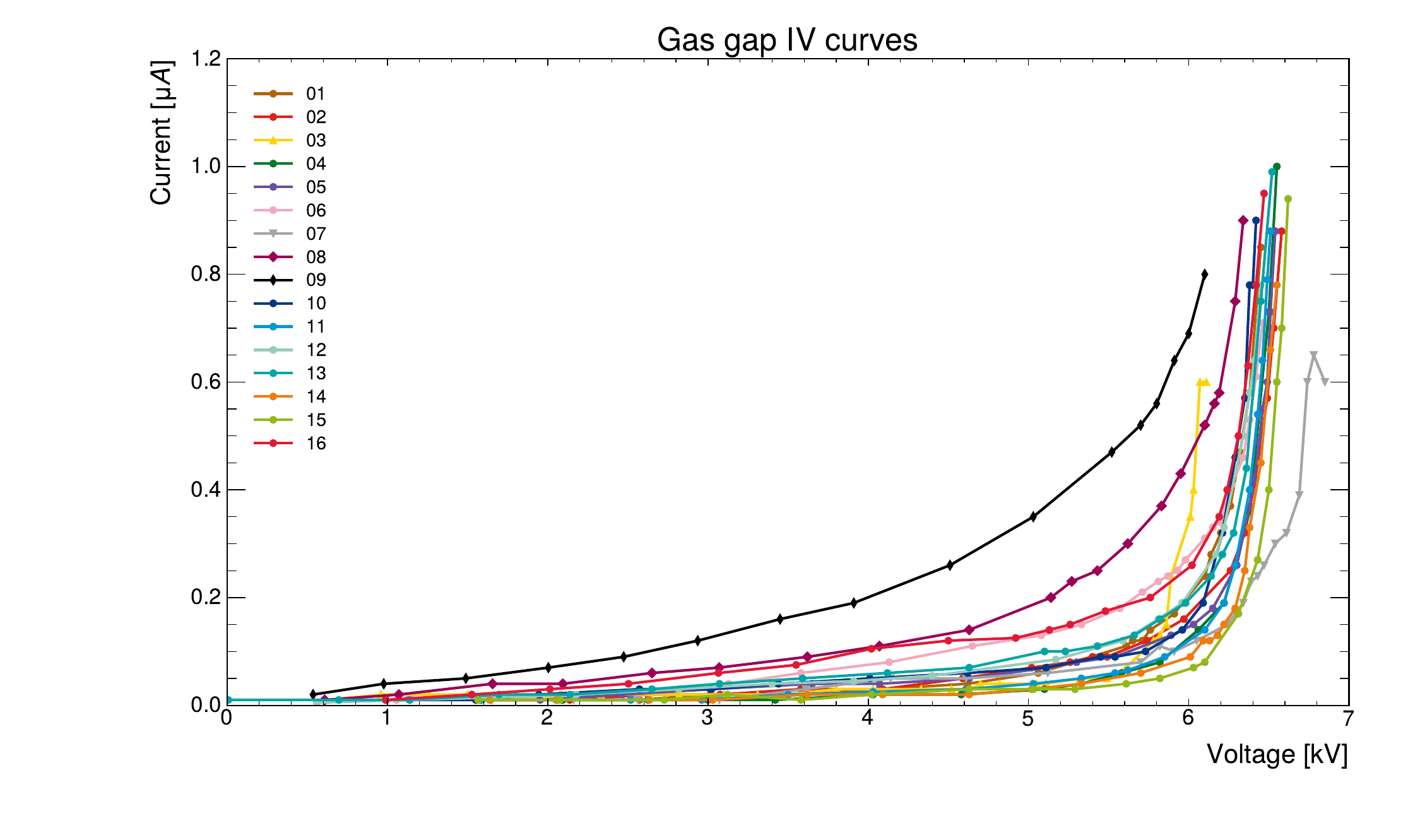}
\caption{The Volt-Ampere (IV) characteristics for the 16 gas gaps received from GTE.} 
\label{fig:Gas_gaps_IV}
\vspace{0.9cm}
\end{figure}

As part of the quality assurance (QA) and quality control (QC) processes, a batch of sixteen gas gaps has been visually inspected to ensure the absence of air bubbles larger than $2\text{--}3$ mm$^{2}$ between the HPL electrodes and the thin Mylar sheet -- typically polyethylene terephthalate (PET) -- glued to the HPL to isolate the graphite coating electrically from the readout strips and to protect it from significant scratches.  The gas gaps are evaluated for ohmic currents measurements, with a quality criterion requiring the total gap current to remain below $1 \mu\text{A}$ at 6 kV, with more discussion provided in Section~\ref{Sec:VI_charateristics}. The measurements are shown in Figure~\ref{fig:Gas_gaps_IV}. Only gas gaps meeting the ohmic current criterion of less than $1\,\mu\text{A}$ are deemed acceptable and used for RPC prototype fabrication.

Each RPC comprises two orthogonal panels (X and Y), approximately 50 cm $\times$ 50 cm in size, for two-dimensional readout. As shown in Figure~\ref{fig:Gas_gaps_stripPanel} (right), each panel is constructed from a Forex sheet, with one side bonded to a PCB strip panel and the other side covered by a copper sheet acting as the ground plate. The panels include 18 copper strips, individually terminated with a 25~$\Omega$ resistor to match the stripline’s characteristic impedance and ensure impedance compatibility with the front-end (FE) electronics. The strips have a pitch of 2.5~cm and are used to read the signals by charged particles. To ensure surface passivation, the copper strips are coated with a thin insulating layer, visible as the green coating in Figure~\ref{fig:Gas_gaps_stripPanel} (right).

\begin{figure}[ht]
\vspace{0.3cm}
\centering
\includegraphics[width=5cm, height=3.4cm]{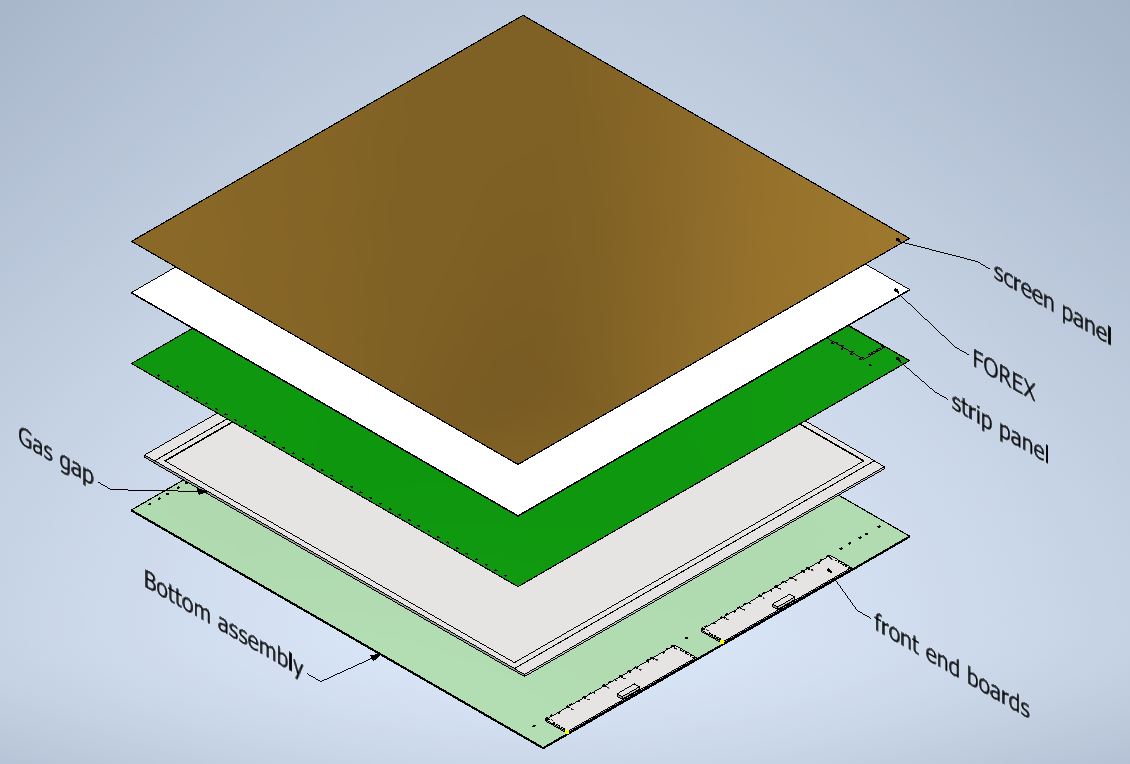}
\includegraphics[width=3cm, height=3.5cm]{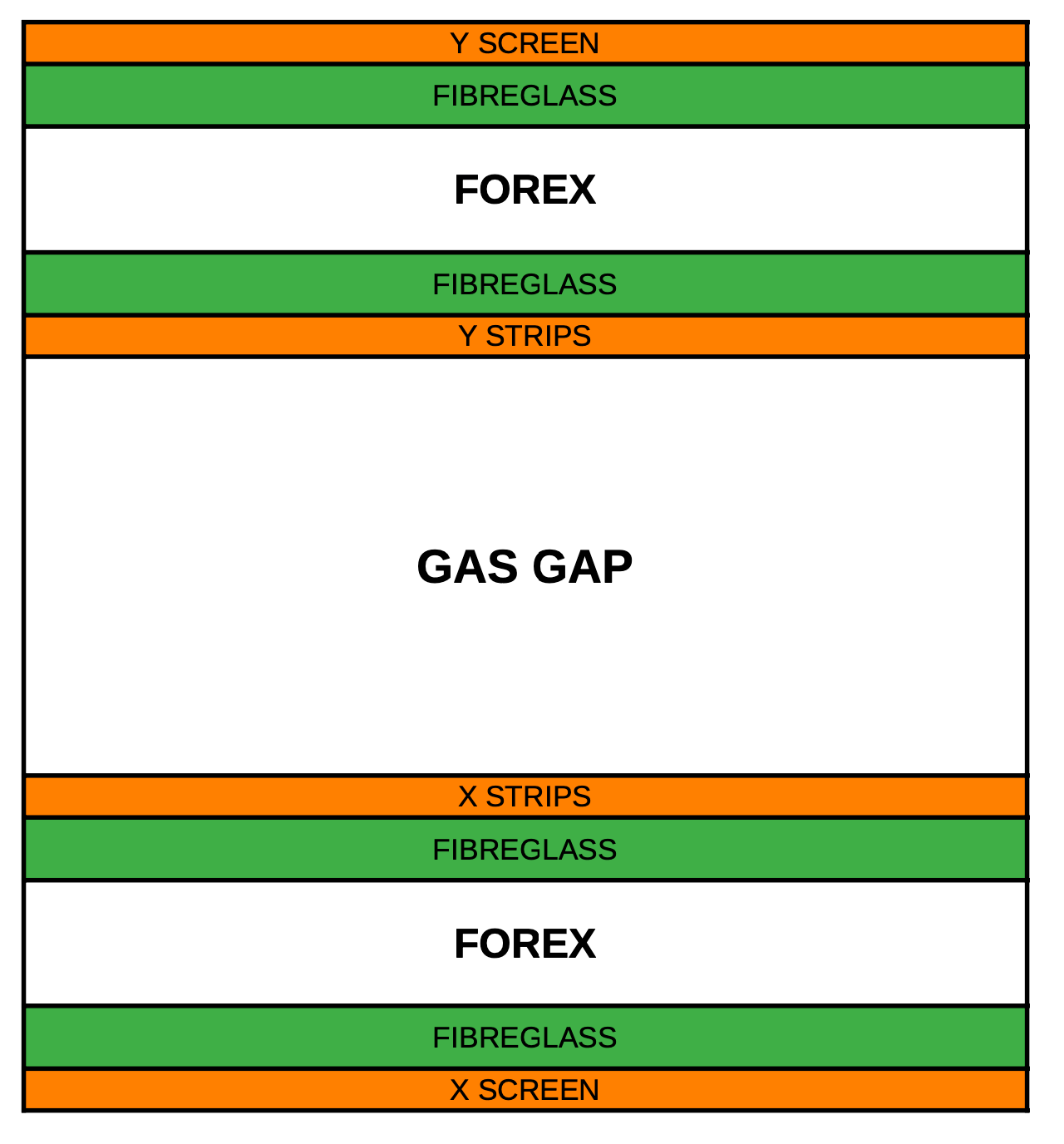}
\vspace{10pt}
\caption{(left) Assembly layout of the prototype RPC, illustrating different components such as the gas gap, strip panels, and the placement of front-end electronics boards on the bottom panel, shown in a pre-assembled state. (right) Cross-sectional view of the RPC structure, detailing its layered components.}
\label{fig:rpc_assembly_layout}
\vspace{0.7cm}
\end{figure}

The gas gap, composed of two HPL electrodes with a graphite coating for high-voltage application and additional components like 1 mm spacers, is enclosed between two orthogonal X and Y strip panels to complete the RPC structure. This assembly is depicted in Figure~\ref{fig:rpc_assembly_layout}. The left side of the figure shows the arrangement of various components, including the placement of the readout electronic FE boards on the bottom panel, while the right side provides a detailed view of the layered structure.

Starting from the bottom, the X-panel consists of a 0.035 mm thick copper ground plane, a 0.36 mm fiberglass layer, a 1 mm Forex sheet, and 0.035 mm thick copper strips. The Y-panel on top has the same structure. Between the X and Y panels lies the gas gap, with a total thickness of approximately 3.5 mm. The gas gap includes 1 mm spacers and two 1.2 mm thick HPL electrodes. Each electrode is made of an HPL sheet, coated on the outer side with an ultra-thin conductive graphite layer to allow high-voltage application. This graphite layer is further insulated from the readout strips by a thin Mylar (PET) sheet.

Each RPC is equipped with two Front-End (FE) boards soldered onto the X-panel and two onto the Y-panel, covering 16 strips in each direction and enabling two-dimensional readout. Out of the 18 strips on each panel, two strips closer to the FE electronics are left unconnected. This is due to the size constraints of the FE boards and to mitigate potential noise interference from the low-voltage supply connected to the FE boards, which could affect the nearby strips.  Each FE board is of the ATLAS BIS78 type~\cite{Pizzimento_2020}, designed to provide the readout of 8 RPC strips or channels.

\subsection{Gas Mixing Setup}
\begin{figure}[ht]
\vspace{0.3cm}
\centering
\includegraphics[width=8.5cm, height=5.8cm]{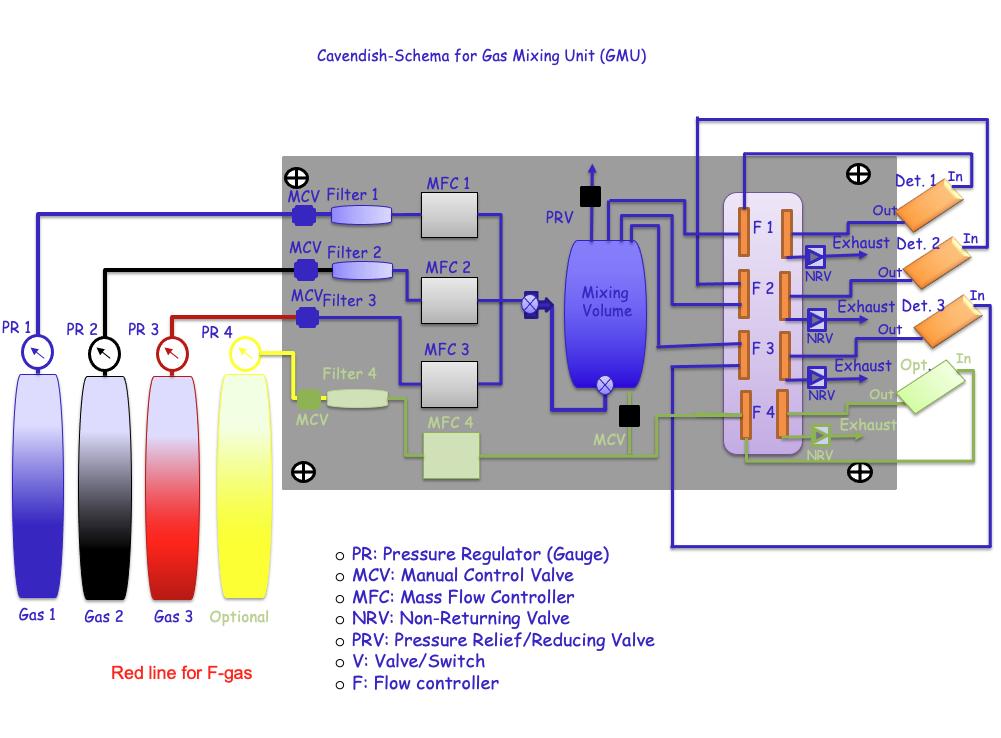}
\vspace{10pt}
\caption{The sketch of a gas mixing setup used in this study. The system comprises four input channels connected to gas bottles, flow controllers, a mixing cylinder, and flowmeters. The components are connected via 6 mm gas tubing with appropriate valves and direct the gas mixture to the RPC chambers for operation.}
\label{fig:Design_GasMixingUnit}
\vspace{0.7cm}
\end{figure}

The gas mixing system used in this study is illustrated in Figure~\ref{fig:Design_GasMixingUnit}. The setup comprises four input channels, each connected to individual gas bottles. The gas flows are regulated using MC-series Mass Flow Controllers (MFCs) from Premier Control Technologies (PCT), Ltd.~\cite{Ref_pct}. The selected gas fractions are then combined within a mixing cylinder to produce the desired mixture, which is subsequently directed to the RPC chambers. The flow rate is controlled within a range of 3 to 4 liters per hour, ensuring stable delivery of the gas mixture for detector operation.

A notable limitation of the MC-series controllers is their incompatibility with Freon-based gases. To address this, a dedicated rotameter is employed for Freon. While this offers sufficient precision, its scale introduces a potential for measurement variation. However, given Freon's dominance as the primary component in the gas mixtures, the impact on overall measurements is minimal. To further enhance precision and streamline operations, the MC-series flow controllers are being upgraded to the latest models -- CODA, again from PCT -- which are compatible with Freon. This improvement will enable more accurate and reliable gas mixture preparation in future studies, thereby reducing measurement uncertainties.

\subsection{Data Acquisition System} \label{Sec:DAQ_setup}
Charged particles, such as muons, passing through RPCs generate avalanches within the gas, inducing signals on the strips. These signals are amplified and shaped by FE electronics, which process and convert them into Low Voltage Differential Signaling (LVDS) format. The LVDS signals are transmitted to custom 6U VME-style trigger boards developed at Cambridge University. These boards implement OR logic, producing a trigger output pulse when one or more inputs are active. This constitutes the first stage of the trigger system. 

In parallel, trigger signals generated by the scintillators (see further Section~\ref{Sec:ScintillatorCalib}) when charged particles pass through them are handled by Silicon Photomultipliers (SiPMs) and undergo initial processing through a discriminator and associated circuitry to convert them into LVDS format. These LVDS signals are then sent to a Field-Programmable Gate Array (FPGA) logic board, which performs coincidence operations among signals from multiple scintillators. The FPGA software enables flexible logic configurations, allowing the selection of signals from individual scintillators or their combinations. This functionality is utilized to generate a trigger signal based on either a single scintillator output or the coincidence of signals from both scintillators, depending on the requirements.

The trigger outputs from the RPC first-stage trigger board and the scintillators, the latter being processed via the FPGA logic board, are aligned with the necessary delays and fed into a second-stage custom 6U VME-style electronics board in the DAQ chain. This board implements AND logic, producing an output pulse only when signals from both systems are simultaneously active. The output pulse is then converted to NIM format and sent to a scaler for counting.

\subsection{Scintillator Trigger Setup}  \label{Sec:ScintillatorTrig}
\begin{figure}[ht]
\vspace{0.3cm}
\centering
\includegraphics[width=5cm, height=6.4cm]{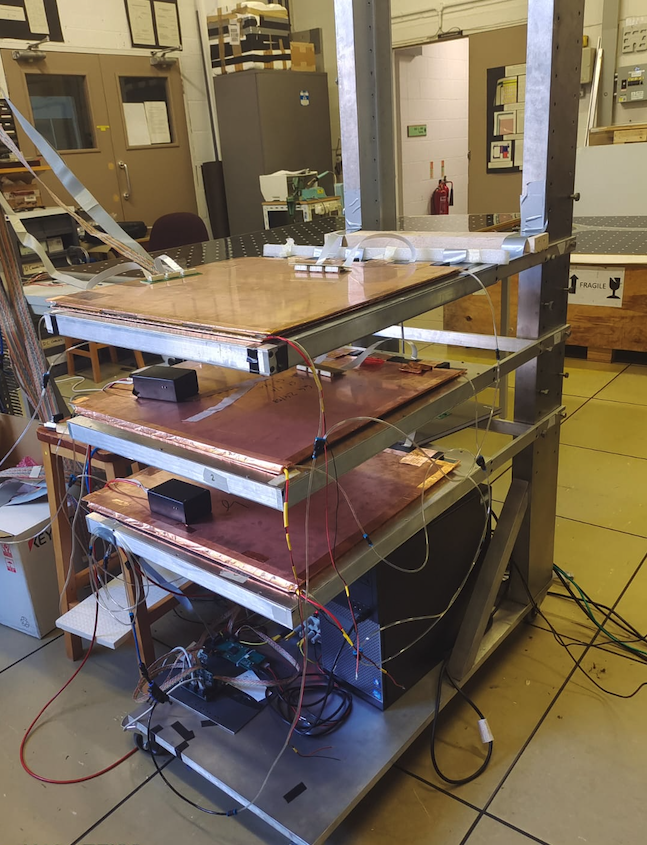}
\vspace{10pt}
\caption{The test setup with three scintillators used for triggering purposes and on top of each scintillator is a prototype RPC.}
\label{fig:rpc_test_setup}
\vspace{0.7cm}
\end{figure}

Three scintillators, each with dimensions of approximately 50 cm $\times$ 50 cm (Figure~\ref{fig:rpc_test_setup}), have been utilised to generate trigger pulses for the RPC efficiency measurements. These scintillators are made of EJ-200 material~\cite{Ref_GScintillator_EJ-204} and paired with two independent Broadcom AFBR-S4N33C013 SiPMs~\cite{Ref_Broadcom}.  The SiPMs are specifically chosen for their spectral sensitivity, which matches the peak emission wavelength of the EJ-200 scintillators, ensuring optimal detection efficiency. 

A dual-SiPM configuration is employed, where two SiPMs simultaneously read the output signal from a single scintillator. This configuration helps to ensure that an output signal is generated only when both SiPMs detect the signal, effectively reducing the likelihood of random noise hits. A custom bias board is used to maintain a stable bias voltage, ensuring the SiPMs operate near their breakdown to optimise their response to the scintillator's light output. Signal discrimination is performed using a comparator board with independent channels, allowing threshold adjustments for each SiPM. Once the scintillator signals exceed the set threshold, as determined by the comparator board, they are transmitted to an FPGA logic board. The FPGA processes these signals and generates the final trigger signal, enabling synchronisation with the RPC muon signals for data acquisition and event recording.

\subsubsection{Calibration of Trigger Setup} \label{Sec:ScintillatorCalib}
The scintillation counters are calibrated to ensure that coincident counts originate from muons rather than dark counts. Calibration involves varying the absolute bias voltage applied to the SiPMs in increments of 2 mV from 600 mV to 790 mV while monitoring the coincidence rate. It is important to note that the SiPMs operate with a negative bias voltage, meaning that increasing the absolute bias voltage corresponds to decreasing the physical voltage applied to the SiPMs. 

\begin{figure}[ht]
\vspace{0.3cm}
\centering
\includegraphics[width=8.5cm, height=5.8cm]{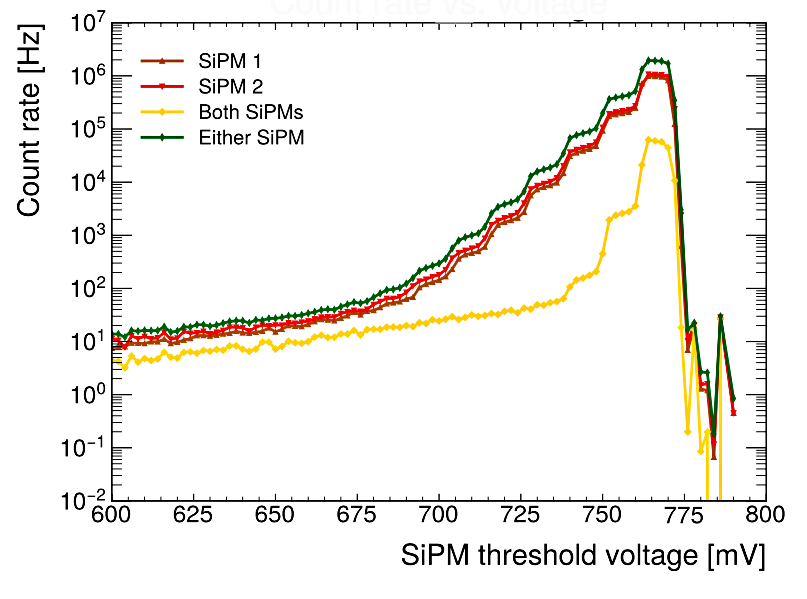}
\vspace{10pt}
\caption{Count rate as a function of threshold voltage for SiPM 1 and SiPM 2 in a given scintillator. `Both' or `Either' SiPMs means logical AND or OR of the two, respectively.}
\label{fig:Scintillator_thesholdScan}
\vspace{0.7cm}
\end{figure}

A threshold voltage scan is shown in Figure~\ref{fig:Scintillator_thesholdScan}, where distinct steps corresponding to discrete numbers of photoelectrons are observed. The count rate, indicated by the curve labelled `Both SiPMs', exhibits a change in gradient around an absolute bias voltage of 730 mV. Below this threshold, the higher effective bias voltage increases the number of photoelectrons required to surpass the threshold, making events more likely to originate from muon-induced excitations. Conversely, fewer photoelectrons are needed to cross the threshold at absolute bias voltages above 730 mV (lower effective bias), allowing a higher rate of dark counts to dominate. To ensure reliable muon detection and minimise dark count interference, the SiPMs are operated at an absolute bias voltage below 730 mV.

\begin{figure}[ht]
\vspace{0.3cm}
\centering
\includegraphics[width=8.5cm, height=5.5cm]{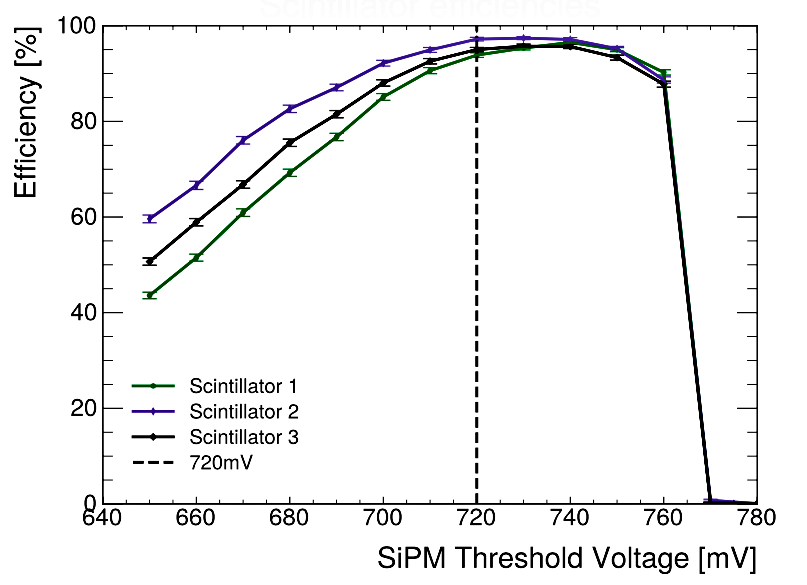}
\vspace{10pt}
\caption{Efficiency as a function of threshold voltage for all three scintillator counters with `OR' logic.}
\label{fig:Scintillator_efficiencies}
\vspace{0.7cm}
\end{figure}

A drop-off in counts is observed at absolute bias voltages above 770 mV. This behaviour can be attributed to the decreasing threshold for signal detection as the effective bias voltage is reduced. At lower thresholds, even small excitations caused by dark noise or thermal fluctuations can generate signals, significantly increasing the rate of photon detections. This elevated photon flux overwhelms the 40 MHz clock interval of the FPGA trigger logic board, which is unable to process signals at the required rate, leading to a drop in observed counts. This highlights the importance of balancing the bias voltage and calibrating it to avoid excessive dark count contributions while maintaining effective muon detection. 

Following calibration, the efficiency of the scintillation counters is measured in a sandwich configuration -- where a test scintillator was placed between two reference scintillators -- as a function of threshold voltage. This approach optimises muon detection while minimising bias toward higher-energy muons. The measured efficiencies, as presented in Figure~\ref{fig:Scintillator_efficiencies}, at 720 mV are 93.9 $\pm$ 0.5\%, 97.2 $\pm$ 0.4\%, and 95.0  $\pm$ 0.3\% for scintillators 1, 2, and 3, respectively. The two best-performing scintillators are selected for subsequent RPC efficiency measurements.

\begin{figure}[ht]
\vspace{0.3cm}
\centering
\includegraphics[width=8.5cm, height=5.8cm]{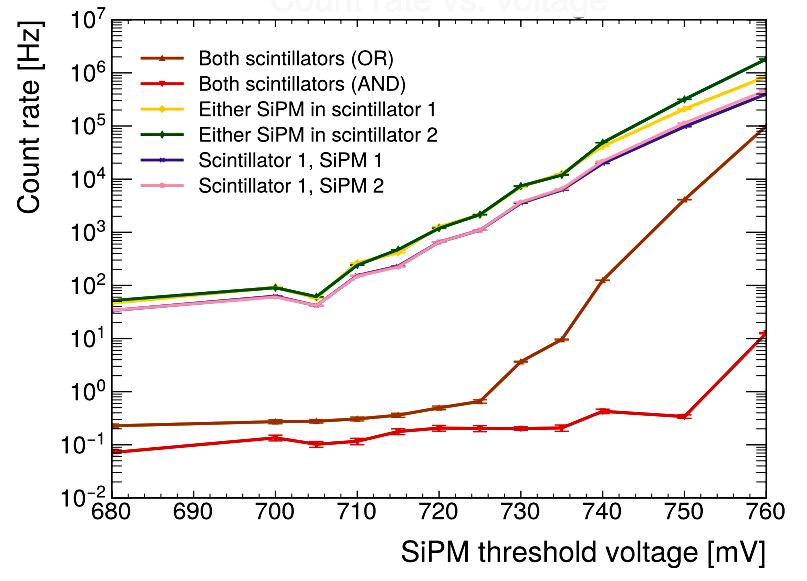}
\vspace{10pt}
\caption{Coincidence rate from two scintillators separated vertically by $\sim$ 30 cm's. The rate curve for ``OR'' logic is when any SiPM in a scintillation is firing and ``AND'' logic is when both SiPMs are firing.}
\label{fig:Scintillator_Coincidence}
\vspace{0.7cm}
\end{figure}

Further, the two scintillators are vertically separated, with their coincidence rate measured across threshold voltages of 680–760 mV, as shown in Figure~\ref{fig:Scintillator_Coincidence}. The increase in coincidences above 730 mV aligns with the transition to the dark count regime, while the plateau below 730 mV indicates correlated muon arrivals from cosmic rays. The rate for the configuration requiring simultaneous signals from both scintillators is slightly lower due to geometric effects, as it depends on muons passing near-equidistant from both SiPMs of a given scintillator to generate the trigger. This favours near-vertical trajectories, minimising the scintillation path length. In comparison, when signals from either scintillator are used, the setup detects muons over a wider range of angles, including those with longer paths through the target scintillator, resulting in slightly higher efficiency. Therefore, to maximise detection efficiency, avoiding angular bias and minimising dark counts, the scintillators are operated when signals from either are registered (i.e., with ``OR'' logic) at an absolute bias voltage of 720 mV. At this threshold, the measured count rate of 37 $\pm$ 1 Hz for both scintillators matches well with the expected sea-level cosmic ray muon flux of approximately 40 Hz~\cite{Ref_CosmicRays}. 

\section{Performance of RPCs with different Gas Mixtures} \label{Sec:Performance_Studies}

\subsection{Volt-Ampere (IV) characteristics} \label{Sec:VI_charateristics}

\begin{figure}[ht]
\vspace{0.3cm}
\centering
\includegraphics[width=7.5cm, height=5cm]{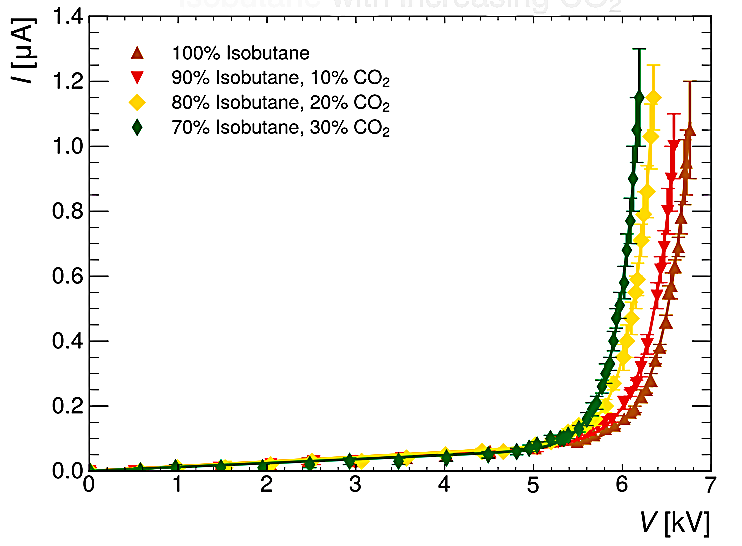}
\vspace{10pt}
\caption{IV characteristics of the RPC with gas gap ID = 8, measured with isobutane and increasing CO$_{2}$ concentration from 0\% to 30\% in steps of 10\%.}

\label{fig:VI_Charaterististics_Isobutane_CO2}
\vspace{0.7cm}
\end{figure}

\begin{figure}[ht]
\vspace{0.3cm}
\centering
\includegraphics[width=7.5cm, height=5cm]{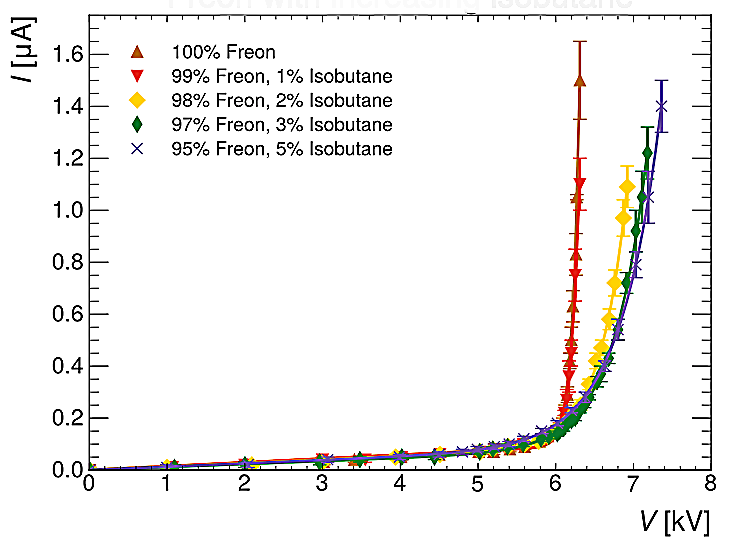}
\includegraphics[width=7.5cm, height=5cm]{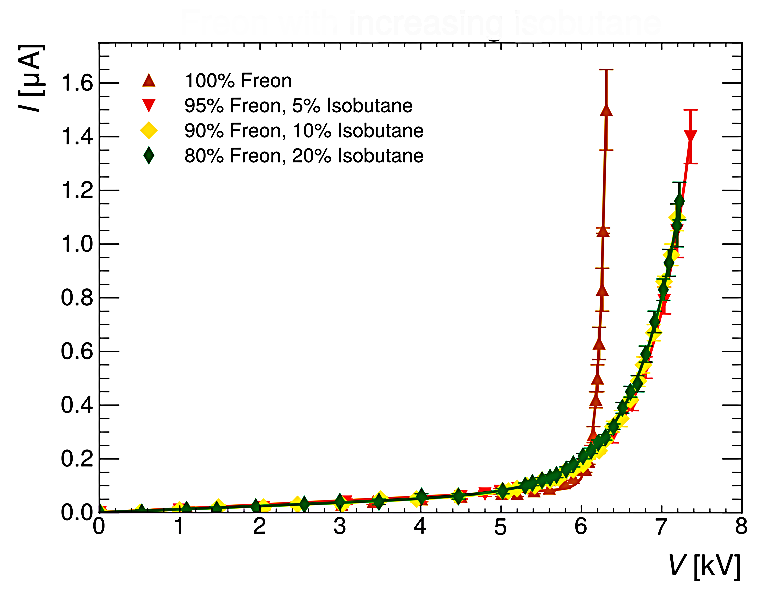}
\vspace{10pt}
\caption{IV curves of the RPC with gas gap ID = 8, measured with Freon (C$_2$H$_2$F$_4$) and increasing concentration of isobutane from (top) 0\% to 5\%, (bottom) up to 20\%.}
\label{fig:VI_Charaterististics_Freon}
\vspace{0.5cm}
\end{figure}

\begin{figure}[ht]
\vspace{0.3cm}
\centering
\includegraphics[width=7.5cm, height=5cm]{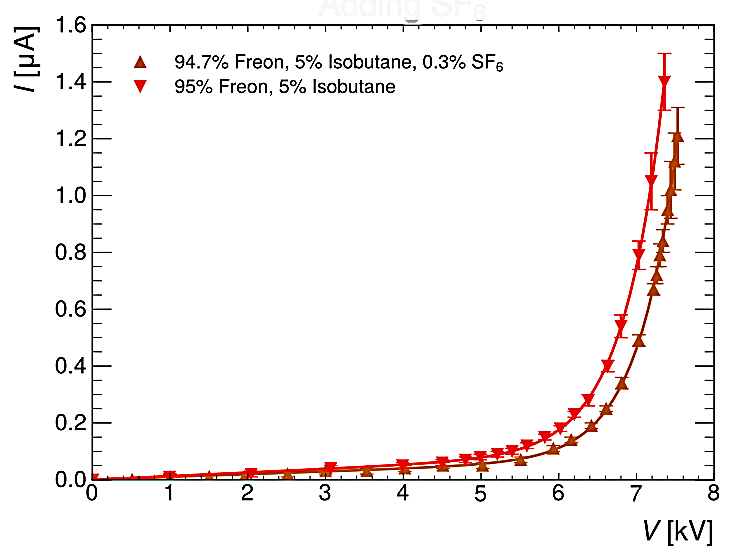}
\vspace{10pt}
\caption{IV curves of the RPC with gas gap ID = 8, for Freon  (C$_2$H$_2$F$_4$) and Isobutane with and without SF$_{6}$.}
\label{fig:VI_Charaterististics_Freon_isobutane_SF6}
\vspace{0.5cm}
\end{figure}

The IV characteristics of gas gaps provide insights into the operational behaviour of RPCs. At low voltages, the electric field within the gas gap is insufficient to initiate avalanche formation. In this regime, the current observed is purely ohmic, arising from conduction through the mechanical structure of the RPC, such as spacers between the electrodes and edge supports around the gas gap. This ohmic current is independent of the gas composition and explains why the IV curves for different gas mixtures generally overlap initially, as in Figure~\ref{fig:VI_Charaterististics_Isobutane_CO2}. As the applied voltage increases, the choice of gas mixture becomes increasingly important. Each gas mixture has a characteristic critical voltage, beyond which the electric field within the RPC is sufficiently strong to ionise gas molecules. The liberated electrons then drift toward the anode under the influence of the applied electric field, initiating avalanches as they traverse the gas. Once this threshold is surpassed, the avalanche process dominates, and the current increases exponentially with voltage due to the rapid multiplication of charge carriers. The total current consists of both the ohmic component and the current resulting from avalanche multiplication, with the latter becoming dominant as the applied voltage increases. As an example, the IV characteristics for pure isobutane with varying concentrations of CO$_2$ are shown in Figure~\ref{fig:VI_Charaterististics_Isobutane_CO2}.

Freon-based IV measurements, Figure~\ref{fig:VI_Charaterististics_Freon}, show a slightly different behaviour when mixed with isobutane. At low concentrations of isobutane (e.g., 1\%), the IV curve remains nearly identical to that of pure Freon (Figure~\ref{fig:VI_Charaterististics_Freon} (top)). However, for isobutane concentrations of 3\% or higher, the IV curves separate distinctly from pure Freon but overlap with one another, indicating that higher concentrations (up to 20\%) have no further impact on the curve's shape (Figure~\ref{fig:VI_Charaterististics_Freon} (bottom)). Interestingly, a 2\% isobutane mixture displays intermediate behaviour, suggesting a transitional regime.

\subsection{Efficiency Measurements}
The performance of an RPC depends critically on the gas mixture used, with each component playing a specific role. Isobutane, for example, acts as a ultraviolet (UV) photon quencher in gas and prevents the formation of secondary avalanches, while Freon and CO$_{2}$ serve as ionising agents that promote avalanche production. SF$_{6}$, when used in small amounts, functions as an electron quencher and limits streamer probability, thus enhancing performance stability. When the applied voltage across the RPC exceeds a critical threshold, the electric field within the gas gap becomes sufficient to ionise gas molecules, initiating avalanche formation as muons pass through. The avalanche current increases exponentially with voltage, as the number of charge carriers grows according to e$^{n}$, where $n$ is the number of carriers created~\cite{Ref_VonEngel}.

RPC efficiency rises sharply as the system enters the avalanche regime, then plateaus after a characteristic voltage rise of a few hundred volts. Optimising efficiency involves identifying the inflection point -- minimum voltage required to reach the plateau -- while minimising streamers and associated ageing effects. However, in some cases, the efficiency curve may exhibit a gradual rise rather than a flat plateau. In such situations, the efficiency at the maximum safe current becomes the important parameter.

\subsubsection{Isobutane-based Mixtures}
\begin{figure}[ht]
\vspace{0.3cm}
\centering
\includegraphics[width=7.5cm, height=5cm]{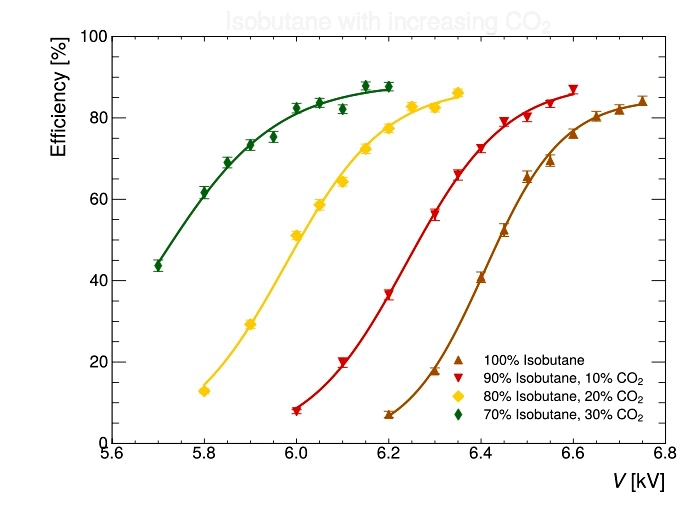}
\vspace{10pt}
\caption{Efficiency for isobutane with increasing CO$_{2}$ fractions, from 0\% to 30\%.}
\label{fig:eff_Isobutane_CO2}
\vspace{0.7cm}
\end{figure}

For initial tests, isobutane is chosen as the primary quenching gas due to its high inflection point and low likelihood of streamer formation, making it straightforward to handle. The isobutane with increasing fractions of CO$_{2}$ yields results as presented in Figure~\ref{fig:eff_Isobutane_CO2}. These measurements reveal a dependence on CO$_{2}$ concentration as the addition of CO$_{2}$ to the isobutane in 10\% steps leads to an increase in currents above the inflection point, as in Figure~\ref{fig:VI_Charaterististics_Isobutane_CO2}, consistent with CO$_{2}$’s lower ionisation potential compared to isobutane. This increase in CO$_{2}$ content also reduced the maximum safe operating voltage, as reflected in the steeper IV curves. 

It should be noted that the results presented in this study have not been corrected for variations in ambient pressure and temperature.

\subsubsection{Freon-based Mixtures}

\begin{figure}[ht]
\vspace{0.3cm}
\centering
\includegraphics[width=7.5cm, height=5cm]{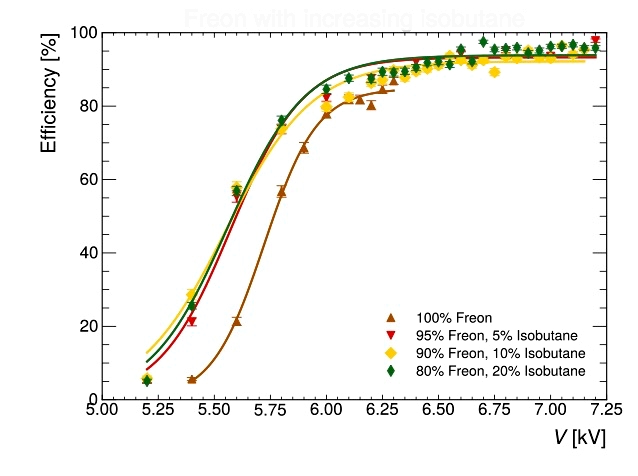}
\includegraphics[width=7.5cm, height=5cm]{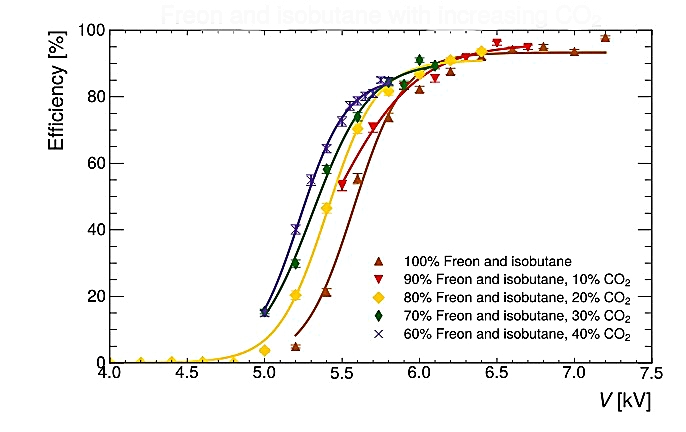}
\vspace{10pt}
\caption{Efficiency for Freon (C$_2$H$_2$F$_4$) (top) with an increasing fraction of isobutane from 0 to 20\% in steps of 5\%, (bottom) with fixed isobutane fraction to 5\% and then increasing fraction of CO$_{2}$ from 0\% to 40\% in steps of 10\%.}
\label{fig:eff_Freon_Isobutane_CO2}
\vspace{0.5cm}
\end{figure}

Efficiency measurements for Freon and its mixtures with isobutane (Figure~\ref{fig:eff_Freon_Isobutane_CO2}) demonstrate relatively better performance, consistent with what is reported in existing literature~\cite{Rigoletti:2875180, Proto:2024bqw}. The study began with a pure Freon followed by a mixture of 95\% Freon and 5\% isobutane, closely resembling the standard RPC gas mixture -- 95.2\% Freon, 4.5\% isobutane, and 0.3\% SF$_{6}$ -- referred to as the CERN mixture in this study, as mentioned previously as well. By varying the isobutane, other fractions such as 10\%, 15\%, and 20\%, have been explored, alongside smaller increments of 1\%, 2\%, and 3\%. It is observed that the efficiency plateaus at 3\% isobutane, with no measurable improvement at higher concentrations. This result suggests the isobutane fraction could be reduced from 5\% to 3\%, offering an additional bit in reducing overall GWP emissions. Additionally, Freon's ionising properties produce sharper inflection points in the IV curves (Figure~\ref{fig:VI_Charaterististics_Freon}) compared to isobutane-based mixtures, aligning with expectations of its rapid transition into the avalanche regime.

Incremental additions of CO$_{2}$ to the Freon-isobutane mixture further reveal critical trends. CO$_{2}$ reduced the critical voltage and steepens the avalanche regime (Figure~\ref{fig:eff_Freon_Isobutane_CO2} (bottom)), while efficiency remains high for CO$_{2}$ concentrations up to 30\%. However, a significant efficiency drop is observed beyond 40\% CO$_{2}$, indicating the upper limit for maintaining detector performance with CO$_{2}$ supplementation.

\begin{figure}[ht]
\vspace{0.3cm}
\centering
\includegraphics[width=7.5cm, height=5cm]{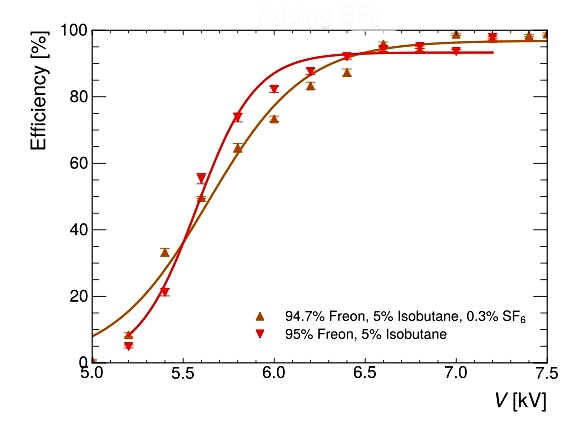}
\includegraphics[width=7.5cm, height=5cm]{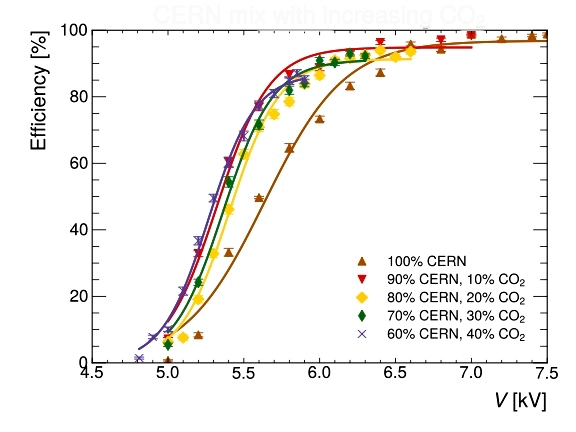}
\vspace{10pt}
\caption{Efficiency: (top) comparison for 95\% Freon (C$_2$H$_2$F$_4$) and 5\% isobutane with the addition of 0.3\% SF$_{6}$, (bottom) standard RPC mixture (95.2\% Freon and 4.5\% of isobutane, and 0.3\% SF$_{6}$) or what has been referred as CERN mixture versus increasing fraction of CO$_{2}$ from 0\% to 40\%, at the cost of Freon.}
\label{fig:eff_CERNmix_SF6_CO2}
\vspace{0.5cm}
\end{figure}

Combining the observations from the 95\% Freon, 5\% isobutane mixture and the standard RPC gas mixture (95.2\% Freon, 4.5\% isobutane, 0.3\% SF$_{6}$), provides further insights into optimising detector performance while reducing environmental impact.

The mixture, incorporating 0.3\% SF$_6$ as a streamer suppressor, exhibits distinct characteristics compared to the 95\% Freon and 5\% isobutane mixture (Figure~\ref{fig:eff_CERNmix_SF6_CO2}~(top)). The addition of SF$_6$ shifts the IV curves (Figure~\ref{fig:VI_Charaterististics_Freon_isobutane_SF6}) by raising the critical voltage and increasing the voltage required to reach a given current in the avalanche regime. This leads to a shallower rise in efficiency beyond the threshold (Figure~\ref{fig:eff_CERNmix_SF6_CO2}), attributable to SF$_6$'s high electronegativity, which suppresses avalanche growth by capturing free electrons. Despite the slower rise, the standard RPC (CERN) mixture achieves a higher overall efficiency due to its improved stability and effective control over avalanche development.

When CO$_{2}$ is added to the CERN mixture, as presented in Figure~\ref{fig:eff_CERNmix_SF6_CO2}, the efficiency trends closely mirror those of the 95\% Freon, 5\% isobutane mix (Figure~\ref{fig:eff_Freon_Isobutane_CO2}). Efficiency remains stable for CO$_{2}$ concentrations up to 40\% but begins to drop off significantly beyond this point while as for 95\% Freon, 5\% isobutane mix, it starts dropping off quickly after adding 30\% of CO$_{2}$. This behaviour suggests that moderate CO$_{2}$ levels maintain the electric field's effectiveness in the avalanche regime while contributing to reduced environmental impact. However, at higher concentrations, CO$_{2}$ begins to overly dilute the mixture's ionisation capabilities, leading to a pronounced decline in efficiency.

\subsubsection*{Discussion:}
The observed behaviours in the measurements can be attributed to the distinct ionisation and quenching properties of the individual gas components. Freon, a strong ionising agent, drives the rapid transition of the gas mixture into the avalanche regime, which is reflected in the sharp inflection points of the IV curves. This property underpins its use as the primary component in standard RPC mixtures. Isobutane acts as a photon quencher, contributing to the stabilisation of avalanche development. The efficiency plateau observed at 3\% isobutane concentration suggests that this level may be sufficient for effective quenching without oversaturating the mixture. This opens the possibility of reducing the isobutane fraction from 5\% to 3\% by increasing the CO$_2$ content, thereby lowering the overall GWP of the mixture. However, ageing studies are necessary to ensure that detector performance and longevity are not compromised.

As a streamer suppressor, SF$_6$ effectively limits ionisation, thereby reducing the likelihood of streamer formation and enhancing the stability of the avalanche regime.  While this shifts the IV curves by raising the critical voltage, it achieves a higher peak efficiency compared to Freon-isobutane mixtures. Additionally, the inclusion of SF$_{6}$ also results in a shallower rise in efficiency beyond the critical voltage, which reflects the balance between enhanced stability and ionisation efficiency.

The addition of CO$_2$ to RPC gas mixtures introduces a trade-off between environmental impact and detector performance. While moderate CO$_2$ levels (up to 30\%) can be incorporated without substantial loss in efficiency, higher concentrations dilute the overall ionisation yield and adversely affect detector operation. This behaviour stems from CO$_2$’s influence on key microscopic parameters, namely the first Townsend coefficient ($\alpha$), which governs the rate of electron multiplication, and the electron attachment coefficient ($\eta$), which represents the likelihood of electron loss. Increasing CO$_2$ concentrations have been shown to reduce $\alpha$~\cite{Auriemma2003Townsend, Sharma1993RPC} and may enhance $\eta$, particularly in the presence of electronegative impurities such as SF$_6$~\cite{Andronic2003Attachment}, leading to increased electron capture and signal attenuation. These effects reduce signal strength and ultimately lower detector efficiency. As such, CO$_2$ shows promise as a low-GWP alternative, but only within carefully optimised concentration limits that preserve acceptable performance.

In our study, 0.3\% of SF$_6$ was included in the gas mixture when higher fractions of CO$_2$ were added to the standard RPC composition. In contrast, the ATLAS collaboration used a higher SF$_6$ concentration of 1\% in their studies when using higher CO$_2$ levels~\cite{Proto:2024bqw}. This difference in SF$_6$ content is expected to influence the avalanche regime and overall detector efficiency. The higher SF$_6$ fraction used by ATLAS likely provides better electron quenching and improved stability against streamers, albeit at the cost of raising the mixture’s critical voltage. Consequently, while the ATLAS mixtures may accommodate higher CO$_2$ concentrations with limited performance degradation, our results show a more pronounced efficiency drop beyond 30\% CO$_2$. This suggests that the lower SF$_{6}$ concentration in our mixtures may exacerbate the dilution of ionisation potential at higher CO$_{2}$ levels, thereby magnifying the trade-off between environmental impact and performance.

\section{Summary and Outlook} \label{sect:Summary}
This study evaluated standard and modified gas mixtures with thin 1 mm gas gap RPCs, establishing benchmarks for assessing their performance and setting the stage for future studies for environmentally friendly alternatives. Future improvements in the experimental setup -- improving trigger calibration and/or employing RPC self-triggering to reduce geometrical misalignments, a refined gas mixing system -- are expected to allow for more systematic measurements. Building on these results, our ultimate aim is to establish an alternative, environmentally friendly gas mixture for the operation of the RPC technology, particularly for the ANUBIS experiment, while ensuring environmental compatibility and contributing to broader applications in high-energy physics. 

Performance measurements with isobutane- and Freon-based mixtures were carried out. For isobutane mixtures, the addition of CO$_{2}$ improved efficiency due to enhanced ionisation effects, whereas Freon-based mixtures experienced a gradual decline in the peak efficiency for CO$_{2}$ concentrations up to 30\%, followed by a more substantial drop at 40\% and beyond. The isobutane fraction in Freon-based mixtures was shown to be reducible from 5\% to 3\%.

The study also underscores the potential to reduce the environmental impact of Freon-based mixtures.  By incorporating CO$_{2}$ up to 30\% offers a practical short- to mid-term pathway to lower GWP emissions without compromising detector performance. The latter adjustment aligns with strategies already explored by the ATLAS experiment during part of Run 3 but there are differences that might affect the performance due to variations in SF$_{6}$ concentration. While ATLAS employed a higher, 1\% SF${_6}$ fraction for CO$_{2}$-rich environments, the 0.3\% used in this study may magnify the trade-off between efficiency and ionisation potential, warranting further investigation.

\section*{Acknowledgements}
We sincerely thank Steve Wotton for his help in setting up and maintaining the scintillator data acquisition (DAQ) system. We are also grateful to our master’s students especially to Patrick Collins for his contributions. Additionally, we would like to thank our dedicated technical support, Richard Shaw and Gaurav Kumar, for their continuous assistance throughout this project.

\bibliographystyle{unsrt}
\bibliography{main,anubis-standard-refs}

\end{document}